\documentclass[%
 aip,
 amsmath,amssymb,
 reprint,
]{revtex4-1}

\usepackage{graphicx}
\usepackage{dcolumn}
\usepackage{bm}

\usepackage[utf8]{inputenc}
\usepackage[T1]{fontenc}
\usepackage{mathptmx}
\usepackage{etoolbox}
\usepackage{hyperref}

\hypersetup{
   colorlinks = true,
   linkcolor  = blue,
   citecolor  = blue,
   urlcolor   = blue
}

\makeatletter
\def\@email#1#2{%
 \endgroup
 \patchcmd{\titleblock@produce}
  {\frontmatter@RRAPformat}
  {\frontmatter@RRAPformat{\produce@RRAP{*#1\href{mailto:#2}{#2}}}\frontmatter@RRAPformat}
  {}{}
}%
\makeatother
\begin{document}

\preprint{AIP/123-QED}

\title{Improving the accuracy of circuit quantization using the electromagnetic properties of superconductors}

\author{Seong Hyeon Park}
  \affiliation{Department of Electrical and Computer Engineering, Seoul National University, Seoul 08826, Republic of Korea}

\author{Gahyun Choi*}
  \affiliation{Center for Superconducting Quantum Computing Systems, Korea Research Institute of Standards and Science, Daejeon 34113, Republic of Korea}

\author{Eunjong Kim*}
  \affiliation{Department of Physics \& Astronomy and Institute of Applied Physics, Seoul National University, Seoul 08826, Republic of Korea}

\author{Gwanyeol Park}
  \affiliation{Quantum Devices Labs, Korea Advanced Nano Fab Center, Suwon, 16229, Republic of Korea}

\author{Jisoo Choi}
  \affiliation{Quantum Design Korea, Seoul, 06272, Republic of Korea}

\author{Jiman Choi}
  \affiliation{Center for Superconducting Quantum Computing Systems, Korea Research Institute of Standards and Science, Daejeon 34113, Republic of Korea}

\author{Yonuk Chong}
  \affiliation{SKKU Adavanced Institute of Nanotechnology, Sungkyunkwan University, Suwon, 16419, Republic of Korea}

\author{Yong-Ho Lee}
  \affiliation{Center for Superconducting Quantum Computing Systems, Korea Research Institute of Standards and Science, Daejeon 34113, Republic of Korea}

\author{Seungyong Hahn*}
  \affiliation{Department of Electrical and Computer Engineering, Seoul National University, Seoul 08826, Republic of Korea}
  \email[Authors to whom correspondence should be addressed: ]{ghchoi@kriss.re.kr, eunjongkim@snu.ac.kr, hahnsy@snu.ac.kr}
\date{\today}

\begin{abstract}
Recent advances in quantum information processing with superconducting qubits have fueled a growing demand for scaling and miniaturizing circuit layouts. 
Despite significant progress, predicting the Hamiltonian of complex circuits remains a challenging task. 
Here, we propose an improved method for quantizing superconducting circuits that incorporates material- and geometry-dependent kinetic inductance. 
Our approach models superconducting films as reactive boundary elements, seamlessly integrating into the conventional circuit quantization framework without adding computational complexity. 
We experimentally validate our method using superconducting devices fabricated with 35 nm-thick disordered niobium films, demonstrating significantly improved accuracy in predicting the Hamiltonian based solely on the device layout and material properties of superconducting films and Josephson junctions. 
Specifically, conventional methods exhibit an average error of 5.4\% in mode frequencies, while our method reduces it to 1.1\%. 
Our method enables systematic studies of superconducting devices with disordered films or compact elements, facilitating precise engineering of superconducting circuits at scale.
\end{abstract}

\maketitle
\section*{Introduction}\label{sec_intro} 
Significant progress in superconducting qubit design~\cite{barend2013coherent, Eun2023shape, gambetta2017investigating, somoroff2023millisecond}, control~\cite{sung2021realization, ding2023high, Li2023error}, and fabrication~\cite{Place2021newmaterial, Wang2022towards, Bal2024systematic} techniques have established superconducting quantum circuits as a promising platform for scalable quantum information processing~\cite{krinner2022realizing, zhao2022realization, google2023suppressing, Sivak2023real, Bravyi2024high-threshold}. 
Realizing high-fidelity quantum gates and extended coherence times in large-scale circuits, however, necessitates stringent fabrication tolerances and yield rates~\cite{Hertzberg2021, morvan2022optimizing}. Generally, an increased number of qubits in a device requires a proportional expansion of auxiliary components, such as readout resonators~\cite{zotova2024control}, couplers~\cite{wei2023compact}, and Purcell filters~\cite{park2024characterization}, which drives the need for more compact circuit layouts. 
This miniaturization introduces challenges, complicating accurate prediction of the system's Hamiltonian.

The Hamiltonian of a superconducting circuit is obtained through canonical quantization, as outlined in circuit quantum electrodynamics~\cite{blais2021circuit}. 
In this framework, the circuit is typically modeled as a lumped-element network, where the flux and charge variables defined at each node are treated as canonically conjugate quantum operators of an oscillator. 
For more complex circuits involving multi-mode, distributed electromagnetic structures, quantization techniques such as black-box quantization (BBQ)~\cite{nigg2012black, solgun2014blackbox, solgun2015multiport, solgun2019simple, solgun2022direct} and energy participation ratio (EPR)~\cite{Minev2021energy, yu2024using} quantization have become widely adopted, demonstrating good agreement between theory and experiment. 
The BBQ method extracts the oscillator properties from the circuit’s driven impedance response, whereas the EPR method employs eigenmode solutions to determine the energy contribution of modes within each element. 
The accurate application of these methods relies on precise electromagnetic characterization of the circuit, usually obtained through classical electromagnetic simulations employing finite-element analysis (FEA).

Although systems comprising superconducting qubits embedded in bulk cavity resonators have been successfully characterized~\cite{solgun2014blackbox, Minev2021energy}, accurately characterizing complex planar circuits remains an outstanding challenge.
In particular, it is known that the use of compact circuit elements~\cite{zhang2023superconducting,park2024characterization} or strongly-disordered superconducting films~\cite{chang2013improved, deng2023titanium} in superconducting quantum devices often leads to notable differences between theoretical models and experimental observations. 
Such discrepancies can be attributed to kinetic inductance present in thin-film superconductors, giving rise to unanticipated downward shifts of resonant frequencies~\cite{park2024characterization, beck2016optimized, frasca2023nbn, porch2005calculation}. 
These kinetic-inductance-induced effects cannot be adequately addressed by conventional quantization methods when only substrate properties are adjusted in FEA simulations~\cite{niepce2020geometric}.
To resolve this challenge, an alternative approach to incorporate distinct electromagnetic behaviors of superconductors needs to be developed.

\begin{figure*}[!t]
  \centering
  \includegraphics[width=0.85\textwidth]{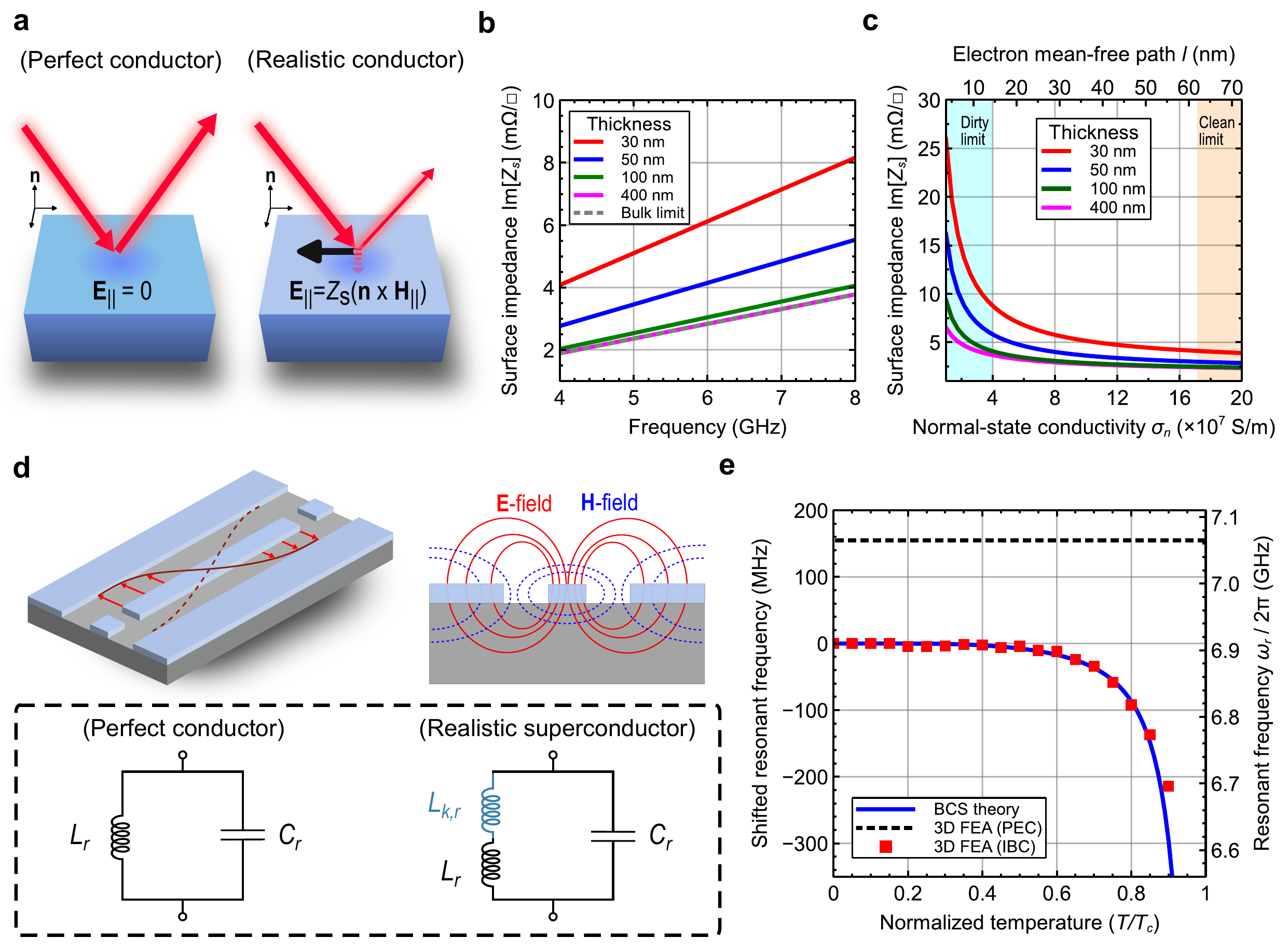} 
  \caption{
  \textbf{Comparison between a perfect conductor and a realistic superconductor.} 
  \textbf{a,} Conceptual illustration of a perfect conductor and a realistic conductor. 
  Obliquely incident electromagnetic wave is totally reflected on the surface of a perfect conductor, with $\textbf{E}_\parallel=0$. In contrast, the incident wave is split into reflected and transmitted waves for a realistic superconductor, resulting in $\textbf{E}_\parallel=Z_s(\textbf{n}\times \textbf{H}_\parallel)$.
  \textbf{b, c,} Calculated surface impedance~$\textrm{Im}[Z_s]$ of a superconducting niobium film with varying thickness $d$ at $T=10$~mK. 
  The niobium parameters are assumed to be $T_c = 9.2$~K, $\lambda_L = 33.3$~nm, $\xi = 39$~nm, and $\Delta_0 = 1.395$~meV consistent with the values in the main text.
  \textbf{b,} Calculated $\textrm{Im}[Z_s]$ as a function of frequency~$\omega$. 
  Here, a fixed $\sigma_n=5.5\times10^7$~S/m is assumed.
  $\textrm{Im}[Z_s]$ approaches to the bulk limit (dashed line), when $d\gg\lambda_\textrm{eff}$.
  \textbf{c,} Calculated $\textrm{Im}[Z_s]$ as a function of $\sigma_n$, evaluated at $\omega/2\pi=7$~GHz.
  $\textrm{Im}[Z_s]$ diverges in the dirty limit (cyan) when $l\gg\xi$, but converges in the clean limit (orange) when $l\ll\xi$.
  \textbf{d,} Top left: Schematic layout of a superconducting half-wavelength CPW resonator. 
  Top right: cross-sectional view of a CPW resonator, where the electric and magnetic field lines are shown as red solid lines and blue dotted lines, respectively.
  Bottom: lumped-element model for the CPW resonator for the cases of a perfect conductor (left) and a realistic superconductor (right). 
  The circuit consists of geometric capacitance $C_r$, inductance $L_r$, and kinetic inductance $L_{k,r}$.
  \textbf{e,} Shift in the resonator frequency as a function of temperature, normalized to $T_c$.
  The shift is calculated with respect to the resonator frequency at zero temperature $\omega_r/2\pi(0~\textrm{K})=6.910~$GHz.
  3D FEA simulations with IBC show close agreement with the BCS thoery~\cite{Watanabe1994kinetic, reagor2013reaching}, while simulation with PEC yields constant resonant frequency $\omega_{r}/2\pi=7.064~$GHz.
  Here, we assume 100 nm-thick niobium film and 500~$\mu$m-thick silicon substrate with a relative permittivity $\varepsilon_r=11.45$.
  }\label{fig_pec_sibc}
\end{figure*}

In this work, we introduce kinetic-inductance-incorporated circuit quantization (KICQ) method, which enables accurate characterization of circuit Hamiltonian by taking into account the electromagnetic properties of superconducting layers.
Here, we approximate thin superconducting films as reactive zero-thickness impedance boundary sheets in full 3D FEA simulation, ensuring that the computational cost remains comparable to conventional methods.
To the best of our knowledge, our method represents a significant departure from prior approaches that characterize superconductors as perfect electric conductors with infinite conductivity.
We test the effectiveness of the KICQ method by characterizing two superconducting devices fabricated from 35 nm-thick niobium film on a sapphire substrate, expected to feature large kinetic inductance.
The KICQ method captures the mode frequencies with an average error 1.1\% and the cross-Kerr frequencies with an average error 11\%, in contrast to the conventional methods exhibiting average errors of 5.4\% and 41\%, respectively. 
The proposed quantization method is straightforward to implement and capable of accurately calculating the circuit Hamiltonian, facilitating precise engineering of large-scale superconducting quantum processors.

\section*{Results}\label{sec_results}
\subsection*{Modeling the electromagnetic behavior of realistic superconductors}\label{sec2_1} 

The key feature of the KICQ method is modeling the realistic electromagnetic behavior of superconductors, in contrast to conventional quantization methods.
For typical electromagnetic problems, conducting layers are often modeled as perfect conductors with infinite conductivity, necessitating that the fields are zero within the conductor. 
The tangential electric field~$\mathbf{E}_\parallel$ at the interface of a perfect conductor is always zero (Fig.~\hyperref[fig_pec_sibc]{1a}, left). 
However, real conductors, including superconductors, exhibit complex conductivity when subjected to oscillating fields.
Therefore, an incident electromagnetic wave can extend into a real conductor, decaying rapidly with distance from the interface. 

While such behavior is often negligible for normal metals, superconductors can exhibit vastly different responses due to their unique electromagnetic properties.
According to the two-fluid model for superconductors~\cite{bardeen1957theory}, paired electrons, or Cooper pairs, carry supercurrent without any loss. 
In contrast, unpaired electrons, known as quasiparticles, can be generated by various mechanisms including thermal excitation, infrared radiation, or cosmic ray, carrying current associated with losses~\cite{Belitsky2006superconducting}.
Furthermore, the finite penetration depth in superconductors results in a non-uniform current distribution, introducing an additional inductive component termed kinetic inductance~\cite{niepce2020geometric}.
These effects can be captured by using the surface impedance $Z_s = R_s + iX_s$, where the real (imaginary) component $R_s=\textrm{Re}[Z_s]$ ($X_s=\textrm{Im}[Z_s]$) represents resistive losses (energy stored) within the superconductor. 
The surface impedance $Z_s$ connects the tangential electric field $\mathbf{E}_\parallel$ to the tangential magnetic field $\mathbf{H}_\parallel$ at the interface, giving an impedance boundary condition (IBC), $\mathbf{E}_\parallel = Z_s (\mathbf{n} \times \mathbf{H}_\parallel)$, that can be utilized in FEA simulations (Fig.~\hyperref[fig_pec_sibc]{1a}, right).  
Here, $\mathbf{n}$ is the normal vector of the interface. Note that the Bardeen–Cooper–Schrieffer (BCS) theory of superconductivity~\cite{bardeen1957theory} predicts $R_s$ to approach zero at sufficiently low temperatures, while $X_s$ converges to a non-vanishing value reflecting the inertial response of Cooper pairs.

In order to calculate $Z_s$, we take into account a parameter set that represents superconducting  material properties, consisting of gap energy $\Delta_0$, critical temperature $T_c$, London penetration depth $\lambda_{L}$, normal-state conductivity $\sigma_n$,  electron mean-free path $l$, and coherence length $\xi$.
These parameters can be obtained through separate calibration experiments or inferred from previous studies.
Then, we calculate the frequency- and temperature-dependent complex conductivity $\sigma_s(\omega, T)=\sigma_1-i\sigma_2$ using Zimmermann's expressions for isotropic superconductors with an arbitrary impurity concentration~\cite{zimmermann1991optical} (see Methods for $\sigma_s$ calculation).
Here, $\sigma_1$~($\sigma_2$) represents the conductivity of quasiparticles (Cooper pairs).
At sufficiently low temperatures $T \approx 0~$K, the effective surface impedance $Z_s$ and effective penetration depth $\lambda_\textrm{eff}$ of a superconductor with thickness~$d$ can be calculated as~\cite{kerr1999surface}
\begin{align}
    Z_s &\approx i\mu_0\omega\lambda_\textrm{eff}\coth{\left(\frac{d}{\lambda_\textrm{eff}}\right)}, \label{eq_Zf} \\
    \lambda_\textrm{eff} &= \sqrt{\frac{1}{\mu_0\omega\sigma_2}}, \label{eq_lambda}
\end{align}
where $\mu_0$ is the permeability of vacuum. 
Since $Z_s$ is purely imaginary and increases proportional to $\omega$, we can rewrite Eq.~\ref{eq_Zf} in terms of the equivalent surface inductance. 
Note that applying an appropriate value of $Z_s$ through the IBC~\cite{U2018modeling, marsic2018nonlinear} enables accurate simulations of the kinetic inductance in superconducting circuits without meshing the complex 3D thin film geometries.
Since the electromagnetic wave propagates as a damped plane wave inside the superconductor with its intrinsic impedance independent of position, the IBC can be utilized to simplify complex volumetric models into computationally manageable surface models~\cite{yuferev2018surface}.

We present the calculated $Z_s$ at $T=10~$mK of a superconducting niobium film for Fig.\hyperref[fig_pec_sibc]{1b,~c} as an example.
Here, the niobium film parameters~\cite{gao2008physics} are assumed to be $T_c=9.2~$K, $\lambda_{L}=33.3~$nm, $\xi=39~$nm, $\Delta_0=1.395~$meV, and $l=20~$nm.
For a fixed $\sigma_n=5.5\times10^7$~S/m, the calculated $\textrm{Im}[Z_s]$ approaches the ``bulk limit'' when the film thickness $d$ significantly exceeds the effective penetration depth ($d \gg \lambda_\textrm{eff}$) (Fig.\hyperref[fig_pec_sibc]{1b}).
The surface impedance is also impacted by the purity of materials often specified by $\sigma_n$ or $l$.
As $\sigma_n$ increases, $\textrm{Im}[Z_s]$ converges to the "clean limit" ($l \gg \xi$)~\cite{Dobrovolskiy2012crossover}, while $\textrm{Im}[Z_s]$ diverges in the "dirty limit" ($l \ll \xi$) (Fig.~\hyperref[fig_pec_sibc]{1c}).
These behaviors highlight the critical dependence of the surface impedance on the purity and thickness of superconductors.

In order to verify the accuracy of FEA simulations with IBC, we reproduce the resonant frequency shifts of a superconducting coplanar waveguide~(CPW) resonator~\cite{oates1991surface, gao2008physics, reagor2013reaching, niepce2020geometric}.
The geometry and the equivalent lumped-element circuits of a CPW resonator are illustrated in Fig.~\hyperref[fig_pec_sibc]{1d}.
Simulations with a perfect electric conductor boundary condition (PEC) demonstrate constant resonator frequency~$\omega_r/2\pi=7.064$~GHz, associated only with geometric capacitance~$C_r$ and inductance~$L_r$ of the CPW resonator.
In contrast, by utilizing IBC with superconducting $Z_s$, the kinetic inductance $L_{k,r}$ in addition to the geometric contribution $L_r$ can be effectively incorporated into the simulation, resulting in temperature-dependent resonator frequency values $\omega_r(T)$ in close agreement with the BCS theory predictions~\cite{Watanabe1994kinetic, reagor2013reaching} (Fig. \hyperref[fig_pec_sibc]{1e}).
Note that the resonator frequency at zero temperature $\omega_r(0~\textrm{K})/2\pi$, expected from both the IBC simulation and the BCS theory, shows about 150~MHz lower value than that from the simulation with PEC, indicating a considerable amount of $L_{k,r}$ present even at $T\approx0~$K.
Further details of the simulation methods are provided in Supplementary Section A.

\subsection*{Principles and quantization procedures of the KICQ method}\label{sec2_3} 

\begin{figure*}[t]
  \centering
  \includegraphics[width=0.82\textwidth]{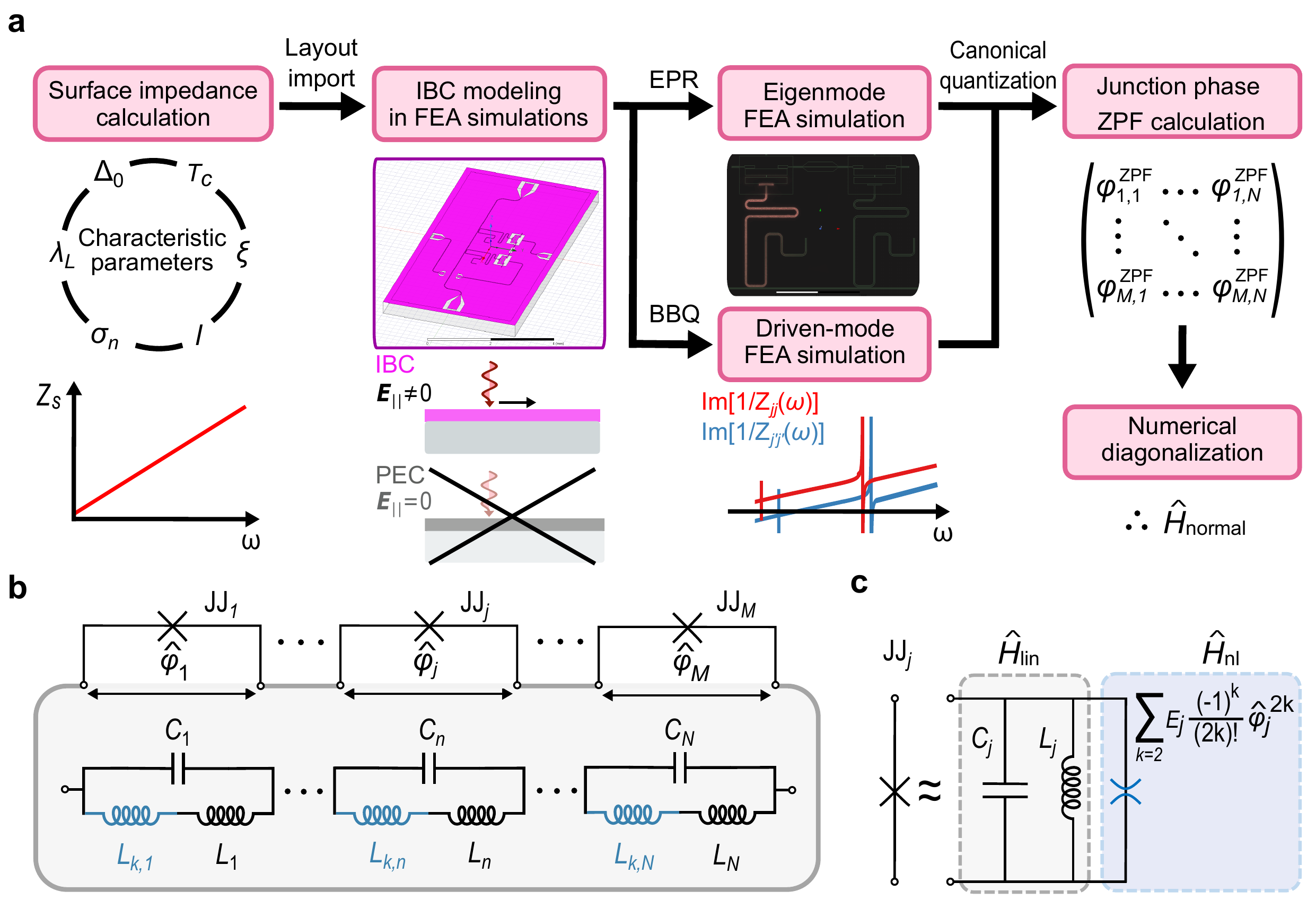}
  \caption{\textbf{Overview of the KICQ method}. 
  \textbf{a,} Overview of the KICQ method. 
  (1) Based on the characteristic material parameters, $Z_s$ of a superconducting film can be calculated. 
  (2) The calculated $Z_s$ as a function of $\omega$ is incorporated through the IBC in FEA simulations. 
  (3) Depending on the choice of the quantization procedure, eigenmode (EPR) or driven-mode (BBQ) FEA simulation is performed to extract the energy participation ratio $p_{j,n}$ or the multi-port impedance matrix $Z_{j,j'}(\omega)$, respectively. 
  (4) From the canonical quantization, the zero-point phase fluctuation $\varphi^\textrm{ZPF}_{j,n}$ can be calculated.
  (5) Finally, the full Hamiltonian can be diagonalized to rewrite it in the normal-mode basis as $\hat{H}_\textrm{normal}$.
  \textbf{b,} Lumped-element circuit diagram with additional kinetic inductance from superconducting layers. 
  The geometric inductance and capacitance are denoted as $L_n$ and $C_n$, respectively. 
  Each frequency-and temperature-dependent kinetic inductance $L_{k,n}$ is connected to $L_n$ in series.
  Here, the Josephson junctions~(JJs) are represented by an `X' symbol, while the electromagnetic environment is modeled as a `black-box' circuit~\cite{nigg2012black,gely2020qucat,Minev2021energy}. 
  \textbf{c,} Equivalent circuit diagram of a Josephson junction element. 
  The junction inductance $L_j$ and capacitance $C_j$ are incorporated into the linear Hamiltonian~$\hat{H}_\textrm{lin}$, while the nonlinear part~$\hat{H}_\textrm{nl}$ is represented by a `$\asymp$' (blue, spider-like symbol) and expressed as a Taylor-expanded series of $E_j(\cos{\hat{\varphi}_j} -1 +\hat{\varphi}^2_j/2!)$. 
  }\label{fig_quantization}
\end{figure*}

The overview of the KICQ method is illustrated in Fig.~\hyperref[fig_quantization]{2a}.
To incorporate the superconducting material properties into conventional quantization methods, the KICQ method utilizes the frequency-dependent $Z_s$ through IBC in FEA simulations.
As a result, the electromagnetic field profiles obtained from eigenmode simulations or the frequency responses from driven-mode simulations are canonically quantized to fomulate the full Hamiltonian of an arbitrary superconducting circuit.
Then, the full Hamiltonian is numerically diagonalized to obtain the Hamiltonian in the normal-mode basis.
Among various quantization methods which utilize electromagnetic FEA simulations~\cite{nigg2012black, solgun2014blackbox, solgun2015multiport, solgun2022direct, dubyna2020inter, Minev2021energy, yu2024using}, the KICQ method employs BBQ~\cite{nigg2012black, solgun2014blackbox, solgun2015multiport, solgun2022direct}, which utilizes the impedance matrix from driven-mode simulations, and EPR~\cite{Minev2021energy, yu2024using}, which calculates the fractional stored electromagnetic energy ratio from eigenmode simulations, to characterize the circuit Hamiltonian.
We believe that our approach can be expanded to lumped-element network theory~\cite{egusquiza2022algebraic, osborne2024symplectic, rodriguez2024geometrical} with appropriate mathematical treatments.

The electromagnetic behaviors of realistic superconductors modeled through the IBC can be integrated into the canonical quantization procedures.
Based on the `black-box' circuit synthesis theorem~\cite{nigg2012black}, the surrounding electromagnetic environments of Josephson junctions can be modeled as a linear sum of   resonant modes.
Since typical superconducting quantum circuits operate at $T \approx 10~$mK and the BCS theory~\cite{bardeen1957theory} predicts negligibly small density of quasiparticles at $T\ll T_c$, we assume the absence of $R_s$ in superconducting films.
We note that the capacitance does not vary significantly with respect to frequency~\cite{guoan2008novel} and kinetic inductance~\cite{Chaudhuri2017broadband}, making it a valid assumption to consider only the additional kinetic inductance.
With the inclusion of additional kinetic inductance from the superconducting layers, an arbitrary black-box circuit can be represented as a lumped-element network illustrated in Fig.~\hyperref[fig_quantization]{2b}.

The full Hamiltonian of Josephson junctions embedded in a black-box circuit can be separated into linear and nonlinear parts.
The Hamiltonian of a superconducting Josephson junction can be approximated by a Taylor-expanded cosine term $\cos{\hat{\varphi}}\approx\sum_n(-1)^{n}\hat{\varphi}^{2n}/(2n!)$, where $\hat{\varphi}$ is the phase difference across the junction, and further separated into the linear and nonlinear parts (Fig.~\hyperref[fig_quantization]{2c}). 
Assuming the reduced Planck's constant $\hbar=1$ for simplicity, the full Hamiltonian of the black-box circuit in Fig.~\hyperref[fig_quantization]{2b} can be expressed as~\cite{gely2020qucat} 
\begin{align}
    \hat{H}_\textrm{full} &= \sum^N_{n=1}\omega_n\hat{a}^\dagger_n\hat{a}_n - \sum^M_{j=1}\sum_{k=2} E_j\frac{(-1)^k}{(2k)!}\hat{\varphi}^{2k}_j, \label{eq_H_rule}\\ 
    \hat{\varphi}_j &= \sum^N_{n=1}\varphi^{\textrm{ZPF}}_{j,n} \left(\hat{a}^\dagger_n+\hat{a}_n\right), \label{eq_H_zpf}
\end{align}
where $\omega_n$ and $E_j$ denote the linear mode frequency and Josephson energy, respectively, while $\hat{a}^\dagger_n$ and $\hat{a}_n$ denote the creation and annihilation operators of mode $n$, respectively.
Here, the cosine potential of Josephson junction can be expanded up to $\hat{\varphi}^4_j$ nonlinearity or higher, depending on the choice of treating the nonlinear terms.
The zero-point phase fluctuation $\varphi^{\textrm{ZPF}}_{j,n}$ of Josephson junction $j$ determines the nonlinear coupling effects to mode $n$.
We highlight that $\varphi^{\textrm{ZPF}}_{j,n}$ is altered significantly if the kinetic inductance~$L_{k,n}$ is comparable to the geometrical inductance $L_n$.

The KICQ method, whether employing the EPR or BBQ method, incorporates $Z_s$ through the IBC. 
However, additional considerations must be taken into account in the implementation of FEA simulations when applying the EPR method to ensure accurate and reliable quantization results.
Since the EPR method employs the total energy stored in the electric and magnetic fields to calculate $\varphi^{\textrm{ZPF}}_{j,n}$, the inductive energy stored in the supercurrent~\cite{beck2016optimized} needs to be considered.
Following similar techniques to those in Ref.~\cite{yu2024using}, the compensated total energy is computed and employed in the KICQ method.
In contrast, the BBQ method does not need any further post-processing of the multi-port admittance matrix from Ref.~\cite{nigg2012black}, aside from incorporating $Z_s$ through  the IBC.
The resulting impedance responses from driven simulations or electromagnetic field distributions from eigenmode simulations are then gathered to compute $\varphi^{\textrm{ZPF}}_{j,n}$ in Eq.~\ref{eq_H_zpf}.
Further details of quantization methods are provided in Supplementary Section B.

As a final step, the Hamiltonian in the normal-mode basis $\hat{H}_\textrm{normal}$ after numerical diagonalization of Eq.~\ref{eq_H_rule} is expressed as
\begin{equation}
    \begin{aligned}
        \hat{H}_\textrm{normal} =& \sum^N_{n=1}{\widetilde{\omega}_n\hat{a}^\dagger_n\hat{a}_n+\frac{\alpha_n}{2}\hat{a}^\dagger_n\hat{a}_n\left(\hat{a}^\dagger_n\hat{a}_n-1\right)}\\ &+ \sum_{n\neq m}{\chi_{m,n}\hat{a}^\dagger_m\hat{a}_m\hat{a}^\dagger_n\hat{a}_n}. \label{eq_H_normal}
    \end{aligned}
\end{equation}
Here, $\widetilde{\omega}_n, \alpha_n$ are the renormalized mode frequency, self-Kerr shift (for weakly anharmonic qubits, anharmonicity) of mode $n$ and $\chi_{m,n}$ is cross-Kerr shift between modes $m$ and $n$, respectively.
For a typical superconducting qubit that is sufficiently detuned from the readout resonator, $\chi$ is often referred to as the dispersive frequency shift.
In this case, additional kinetic inductance in a resonator is expected to reduce its mode frequency, thereby affecting the dispersive frequency shift through the modified qubit-resonator detuning~\cite{blais2021circuit} $\Delta_{QR}=\widetilde{\omega}_Q-\widetilde{\omega}_R$, as $\chi_{QR} \propto 1/(\Delta_{QR})^2$.
However, due to the large inductance of the Josephson junction compared to the added kinetic inductance, the frequency and the anharmonicity of transmon qubits are expected to remain nearly unchanged.
Such effect cannot be reproduced by simply modifying the substrate's relative permittivity $\epsilon_r$ in FEA simulations, as this adjustment affects not only the resonator frequency but also the qubit transition frequency and anharmonicity via changes in geometric capacitance.

\subsection*{Experimental Verification}\label{sec3_1} 
\begin{figure*}[t]
    \centering
    \includegraphics[width=0.9\textwidth]{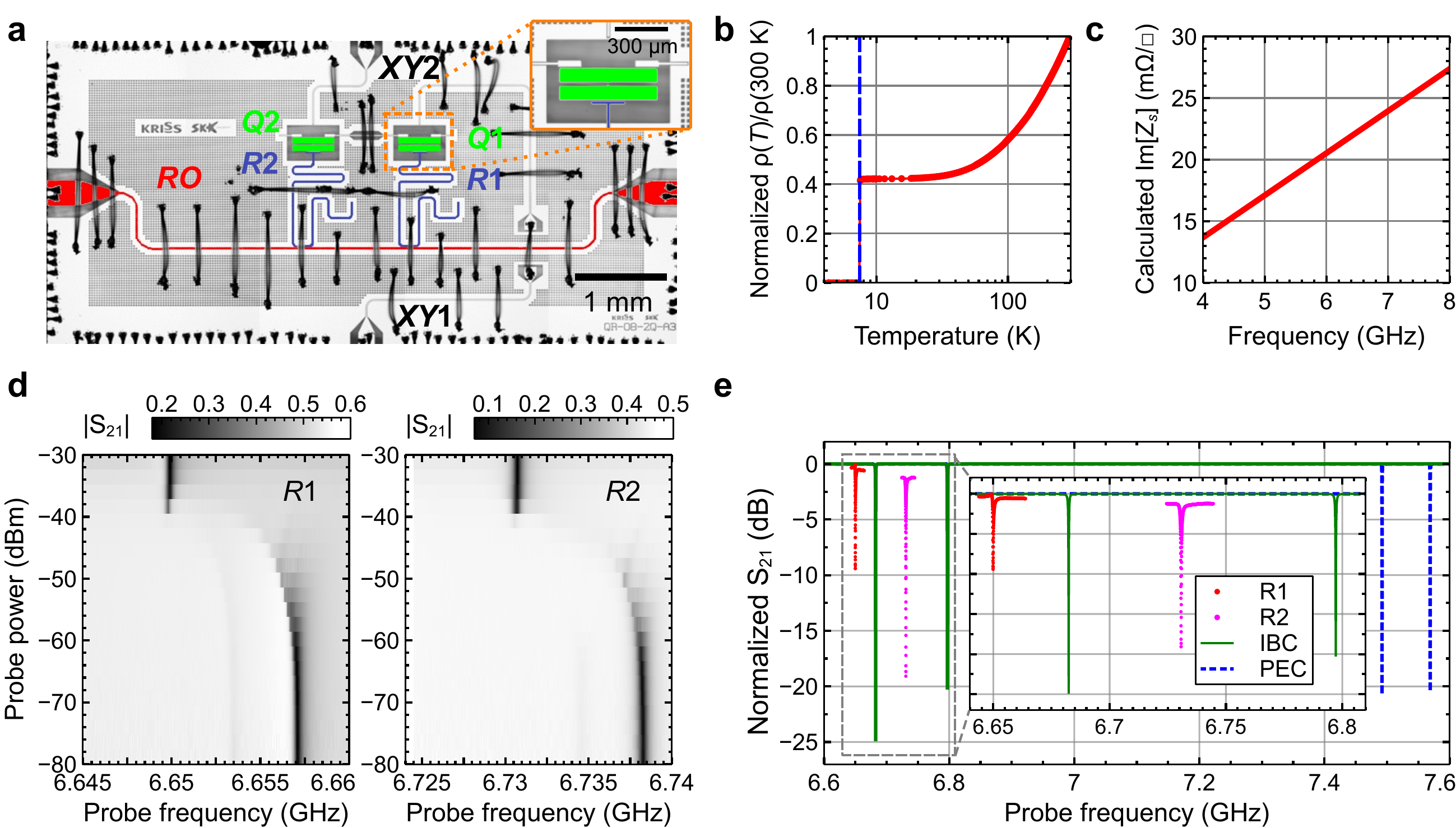} 
    \caption{
    \textbf{Characterization of a two-qubit device with highly-disordered thin film}. 
    \textbf{a,} False-colored optical micrograph of a planar two-qubit device. 
    Fixed-frequency transmon qubits~(green, \textit{Q}1 and \textit{Q}2), capacitively coupled to CPW readout resonators~(blue, \textit{R}1 and \textit{R}2), have identical geometric features.
    The inset (orange) shows the enlarged view of \textit{Q}1 geometry. 
    The readout resonators are capacitively coupled to the readout (red, \textit{RO}) transmission line. 
    \textbf{b,} Measured film resistivity $\rho(T)$ normalized to the value at $T=300~$K. 
    Blue dash line indicates $T_c=7.47~$K. 
    \textbf{c,} Calculated $\textrm{Im}[Z_s]$ at frequencies between 4~GHz and 8~GHz. 
    \textbf{d,} Transmission spectrum of \textit{R}1 and \textit{R}2 as a function of probe frequency and power. 
    Here, the probe power is defined as the output power from the network analyzer. 
    At a high probe power above -37~dBm, punch-out of the readout resonators to the bare resonator frequencies is observed.
    \textbf{e,} Measured and simulated transmission spectra at a high probe power. 
    The transmissions of \textit{R}1 (red circles) and \textit{R}2 (magenta circles) are measured with a probe power of -30 dBm and normalized to the maximum value. 
    Driven-mode simulations are conducted using superconducting material properties (IBC, green solid line) and PEC assumption (blue dashed line).
    The inset provides a zoomed-in view of the resonant frequencies for \textit{R}1 and \textit{R}2.
    }\label{fig_chip}
\end{figure*}

We prepared a two-qubit device, fabricated with 35 nm-thick niobium film on a 550 $\mu$m-thick c-plane sapphire substrate (Fig.~\hyperref[fig_chip]{3a}).
The Josephson junctions were fabricated by a standard double-angle evaporation Al/AlO$_x$/Al process, following techniques similar to those in Ref.~\cite{park2020observation}.
For detailed information regarding the device fabrication, refer to the Methods section.

The niobium thin film used in the two-qubit device was characterized using a standard four-point probe method to measure direct-current resistivity~$\rho(T)$ as a function of temperature $T$ (Fig.~\hyperref[fig_chip]{3b}).
From this measurement, we determine the critical temperature of $T_c=7.47~$K and the residual resistivity ratio~(RRR) to be $\rho(300~\textrm{K})/\rho(T_c)=2.30~(\sigma_n=1/\rho(T_c)\approx1.15\times 10^7$ S/m), both consistent with reported values for disordered niobium films~\cite{gubin2005dependence, niepce2020geometric}. 
These observations indicate the presence of strong disorder within the niobium film, presenting challenges for accurate modeling using conventional circuit quantization methods.
Based on the measured film parameters, we calculate the complex conductivity $\sigma_s$ and the corresponding $Z_s$ using Eq.~\ref{eq_Zf}.
In contrast to clean and thick superconductors, the calculated imaginary part of $Z_s$ (Fig.~\hyperref[fig_chip]{3c}) shows values large enough to study the material-property-induced variations in the circuit Hamiltonian parameters.

We measure the transmission spectrum $\textrm{S}_{21}$ of CPW readout resonators \textit{R}1 and \textit{R}2, capacitively coupled to qubits \textit{Q}1 and \textit{Q}2, respectively, as a function of probe frequency and power (Fig.~\hyperref[fig_chip]{3d}).
When the number of photons in the readout resonator exceeds a certain threshold, the resonance features shift toward the bare readout resonator frequencies of $\omega_{R1, \textrm{bare}}/2\pi=6.6499~$GHz and $\omega_{R2, \textrm{bare}}/2\pi=6.7307~$GHz.
Since the CPW resonators were designed with target resonant frequencies of 7.5 GHz~(physical length approximately 4.2 mm) and target spacing of 100 MHz, the difference between the measured and designed CPW resonator frequencies indicates a significant contribution of kinetic inductance from the niobium film.
To verify this, we perform FEA simulations with and without incorporating the superconducting material properties (Fig.~\hyperref[fig_chip]{3e}).
Simulations with PEC exhibit discrepancies of approximately 800 MHz, whereas notable agreement is observed between the measurements and the IBC simulations, with only 30 MHz of discrepancies.

The measured and simulated Hamiltonian parameters of the two-qubit device are summarized in Table~\ref{table_2Q_chip}.
The qubit frequency $\omega_Q$, readout resonator frequency $\omega_R$, qubit anharmonicity $\alpha$, and dispersive shift frequency $\chi_{Q,R}$ between $Q$ and $R$ are calculated by the conventional and KICQ methods.
Both the conventional and KICQ methods show good agreement between the measured and simulated qubit parameters, $\omega_Q$ and $\alpha$.
In the absence of additional kinetic inductance~(PEC in Table~\ref{table_2Q_chip}), conventional EPR and BBQ methods predict that $\omega_R$ is approximately 12\% larger, while $\chi_{Q,R}$ are about 40\% smaller than the measured values.
As described in the previous section, simulations with PEC often lead to a significant misinterpretation of mode frequencies due to the neglect of kinetic inductance in superconducting films. 
In contrast, incorporating superconducting material properties into the quantization method~(KICQ in Table~\ref{table_2Q_chip}) improves the accuracy in predicting $\omega_R$ and $\chi_{Q,R}$ by at least five-fold.
This result indicates that the KICQ method effectively accounts for variations in Hamiltonian parameters induced by kinetic inductance.
Detailed discussion on the modeling of a transmon qubit is provided in Supplementary Section C, with the comparison results summarized in Supplementary Table S1.

\begin{table*}[!t]
\caption{\textbf{Summary of the measured and simulated characteristic Hamiltonian parameters for the two-qubit device.} The simulation errors in parentheses are defined as relative errors normalized to the measured values to quantify the agreement of each quantization method. Josephson junctions are replaced with a lumped-element circuit with a inductor $L_J$ and a capacitor $C_J$ in parallel. The dispersive shifts are the total difference in the readout resonator's frequency depending on the qubit's state defined as $(\omega_{|11\rangle}-\omega_{|10\rangle})-(\omega_{|01\rangle}-\omega_{|00\rangle})$, where ${|q,r\rangle}$ denotes the eigenstate of the qubit in state $q$ and the resonator in state $r$.}\label{table_2Q_chip}
\begin{tabular*}{0.95\textwidth}{@{\extracolsep\fill}l c c c c c}
\toprule
Parameters & Measurement & EPR & EPR & BBQ & BBQ \\
& & (PEC) & (KICQ) & (PEC) & (KICQ) \\
\hline
$Q_1$ frequency & 4.7595 & 4.8131 & 4.8113 & 4.7835 & 4.7944\\
$\omega_{Q1}/2\pi$ (GHz) && (1.13\%) & (1.09\%) & (0.50\%) & (0.73\%) \\

$Q_1$ anharmonicity& $-342.3$ & $-343.2$ & $-342.6$ & $-340.6$ & $-339.9$\\
$\alpha_1/2\pi$ (MHz) && (0.26\%) & (0.09\%) & (0.50\%) & (0.70\%) \\

$R_1$ frequency & 6.6571 & 7.4959 & 6.6909 & 7.5107 & 6.7109 \\
$\omega_{R1}/2\pi$ (GHz) && (12.60\%) & (0.51\%) & (12.82\%) & (0.81\%) \\

$Q_1$-$R_1$ dispersive shift & $-3.6$ & $-2.19$ & $-3.43$ & $-2.37$ & $-3.84$\\
$\chi_{Q1,R1}/2\pi$ (MHz) && (39.06\%) & (4.72\%) & (34.17\%) & (6.67\%) \\
\hline
$Q_2$ frequency & 4.8342 & 4.8668 & 4.8654 & 4.8458& 4.8560\\
$\omega_{Q2}/2\pi$ (GHz) && (0.67\%) & (0.65\%) & (0.24\%) & (0.45\%) \\

$Q_2$ anharmonicity & $-344.0$ & $-345.9$ & $-344.6$ & $-342.1$ & $-341.7$\\
$\alpha_2/2\pi$ (MHz) && (0.55\%) & (0.17\%) & (0.55\%) & (0.67\%) \\

$R_2$ frequency & 6.7383 & 7.5796 & 6.8107 & 7.6012 & 6.7971\\
$\omega_{R2}/2\pi$ (GHz) && (12.49\%) & (1.07\%) & (12.81\%) & (0.87\%) \\

$Q_2$-$R_2$ dispersive shift & $-3.7$ & $-2.19$ & $-3.42$ & $-2.32$ & $-3.68$\\
$\chi_{Q2,R2}/2\pi$ (MHz) && (40.78\%) & (7.57\%) & (37.30\%) & (0.54\%) \\
\toprule
\end{tabular*}
\end{table*}

\begin{figure*}[!t]
    \centering
    \includegraphics[width=0.9\textwidth]{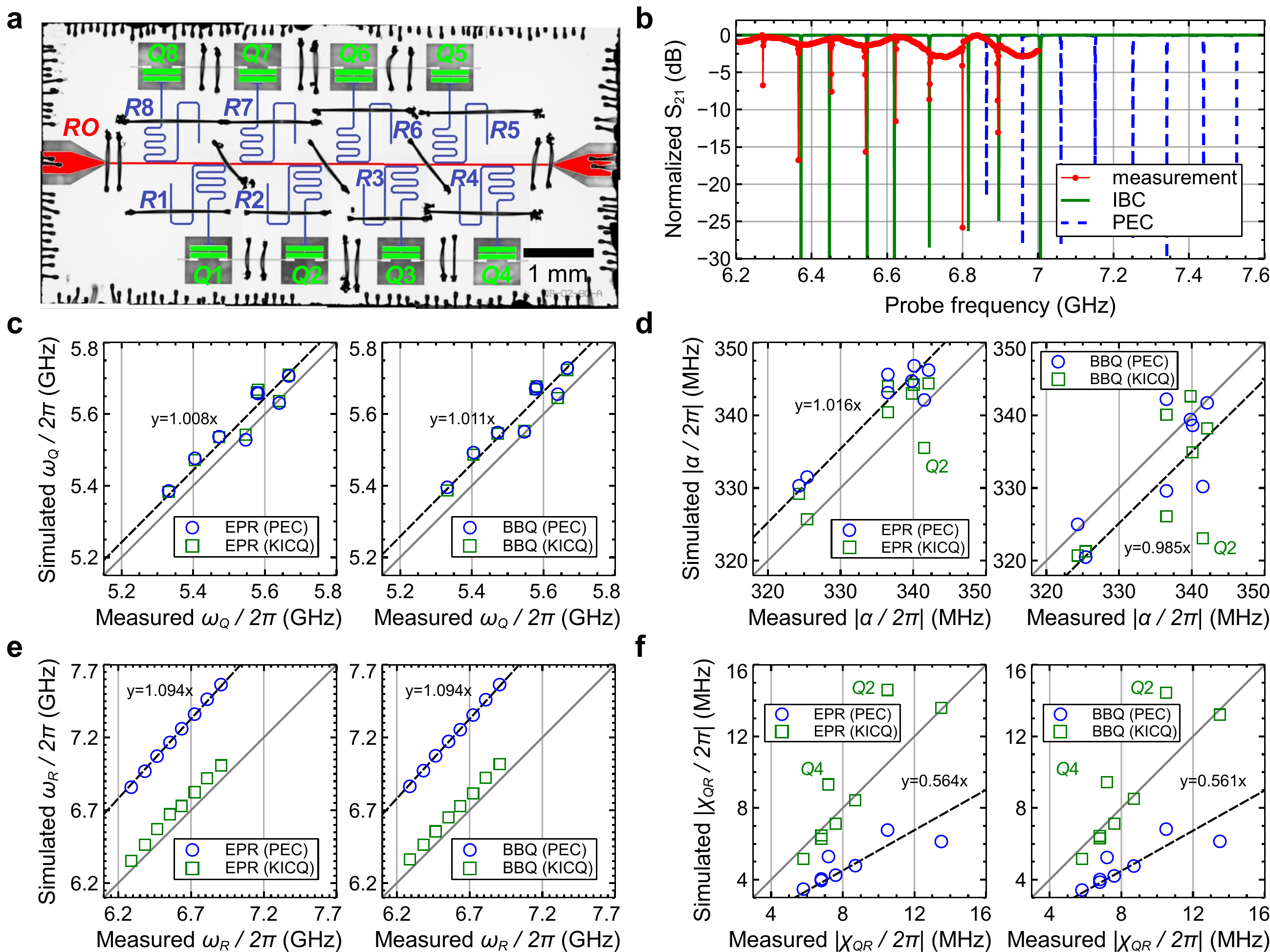} 
    \caption{
    \textbf{Benchmarking the KICQ method using an eight-qubit device}. 
    \textbf{a,} False-colored optical micrograph of a planar eight-qubit device. 
    Similar to the two-qubit device in Fig.~\hyperref[fig_chip]{3a}, fixed-frequency transmon qubits~(green, \textit{Q}1 - \textit{Q}8) are coupled to CPW readout resonators~(blue, \textit{R}1 - \textit{R}8). 
    \textbf{b,} Measured and simulated transmission spectrum $\textrm{S}_{21}$. 
    Measurements are indicated by red circles, while simulations with PEC and IBC are indicated by blue dashed line and green solid line, respectively. 
    Here, measured $\textrm{S}_{21}$ are normalized to its maximum value. 
    \textbf{c-f,} Comparison between conventional methods and the KICQ method. 
    blue circle symbols indicate predicted values by conventional EPR or BBQ methods, while green rectangular symbols indicate predicted values by the KICQ method. 
    The solid gray lines indicate the ideal agreement between simulation and measurement ($y = x$), while the dashed black lines correspond to the fitted functions of conventional methods. 
    \textbf{c,} The measured and the simulated qubit frequency~$\omega_Q$.
    \textbf{d,} Magnitudes of the measured and the simulated anharmonicity~$|\alpha|$.
    \textbf{e,} The measured and the simulated readout resonator frequency~$\omega_R$.
    \textbf{f,} Magnitudes of the measured and the simulated dispersive shift~$|\chi_{QR}|$.
    In panels \textbf{d} and \textbf{f}, outliers are indicated with the corresponding qubit labels.
    }\label{fig_8Q_chip}
\end{figure*}

The eight-qubit device (Fig.~\hyperref[fig_8Q_chip]{4a}) was fabricated with 35 nm-thick niobium film on a sapphire substrate, prepared on a separate wafer from the two-qubit device. 
From the four-point probe measurement, $T_c$ of the niobium film is determined to be 8.24~K, with $\textrm{RRR}=\rho(300~\textrm{K})/\rho(T_c)=3.84$. 
These results indicate that the niobium film in the eight-qubit device has lower disorder compared to that in the two-qubit device, leading to smaller $Z_s$ values.

The measured and simulated $\textrm{S}_{21}$ of the eight-qubit device over a broad frequency range are plotted in Fig.~\hyperref[fig_8Q_chip]{4b}.
The readout resonators are quarter-wavelength CPW resonators, designed with target resonant frequencies ranging from 6.8 GHz to 7.5 GHz, spaced at 100 MHz intervals.
The measured resonator frequencies, however, are in range from 6.3 GHz to 6.9 GHz, showing the uniform downward shift of approximately 600 MHz.
Such discrepancies indicate a significant contribution of the kinetic inductance to the mode frequencies.
In contrast, simulations with IBC achieves a notable correction to the simulated $\textrm{S}_{21}$, resulting in discrepancies of only 100 MHz.
We note that further correction can be achieved by reflecting larger values of $Z_s$, e.g., by choosing a larger value of $\lambda_L$.

We perform the conventional and KICQ methods to the eight-qubit device, consisting of eight fixed-frequency qubits and eight readout resonators.
The qubits have identical geometric features with target anharmonicity of approximately $-340~$MHz.
$L_J$ and $C_J$ of each transmon qubit are extracted using methodologies similar to those employed in the two-qubit device simulations. 
Further details on the eight-qubit device simulations are provided in Supplementary Section D.

The simulated qubit parameters, including $\omega_Q$~(Fig.~\hyperref[fig_8Q_chip]{4c}) and $\alpha$~(Fig.~\hyperref[fig_8Q_chip]{4d}), show differences less than 2\% between conventional methods and the KICQ method.
With appropriate assumptions for $L_J$ and $C_J$, both conventional methods and the KICQ method calculate the qubit parameters that closely align with the measured values for the eight-qubit device.
The BBQ method predicts slightly smaller values for $\alpha$ compared to the EPR method, although the overall absolute errors remain nearly identical.
Summary of the measurement and simulation results of the eight-qubit device are provided in Supplementary Table~S2.

The conventional methods demonstrate huge discrepancies in predicting $\omega_R$~(Fig.~\hyperref[fig_8Q_chip]{4e}) and $\chi_{Q,R}$~(Fig.~\hyperref[fig_8Q_chip]{4f}), whereas the KICQ method demonstrates better agreement with the measurements.
However, there are two outliers ($\chi_{Q2,R2}$ and $\chi_{Q4,R4}$) that the KICQ method predicts the overestimated $\chi_{Q,R}$ values. 
The KICQ method accurately predicts the detuning frequencies of the outliers in close alignment with the measurements but predicts corresponding $\chi_{Q,R}$ values with about 40\% larger than the measurements.
We attribute these results to imperfect assumptions about the junction parameters, as indicated by larger errors in $\omega_Q$ compared to other qubits. 
Despite several outlying data, the KICQ method demonstrates significantly improved accuracies in $\omega_R$ and $\chi_{Q,R}$ for superconducting circuits.
Overall, the KICQ method demonstrates average errors of 0.9\% for $\omega_Q$, 1.3\% for $\alpha$, 1.3\% for $\omega_R$, and 11\% for $\chi_{Q,R}$, compared to conventional methods, which demonstrate average errors of 1.0\%, 1.2\%, 9.7\%, and 41\%, respectively.

\section*{Discussion}\label{sec_discussion} 
The proposed circuit quantization method allows for accurate predictions of mode frequencies and Kerr shift frequencies of superconducting devices without additional computation cost.
Our findings indicate that kinetic inductance is crucial in determining Hamiltonian parameters.
Recent studies on aluminum~\cite{li2023experimentally} and tantalum~\cite{varmazis1974inductive}, popular choices for fabricating quantum circuits, have revealed that the kinetic inductance of well-known clean superconducting films cannot be ignored, giving rise to resonator frequency shifts of at least tens of MHz.
While partial etching of ground layers~\cite{li2023experimentally} and metal encapsulation~\cite{chang2024eliminating} can control the resonant frequency shifts caused by kinetic inductance, we note that incorporating the electromagnetic properties of clean superconductors can lead to a more accurate prediction of the circuit Hamiltonian.

This method can be readily employed in other applications of superconducting circuits based on disordered superconducting films, such as parametric amplifiers and microwave kinetic inductance detectors.
In particular, kinetic inductance traveling wave parametric amplifiers~(KI-TWPAs) utilizing superconducting transmission lines~\cite{hoeum2012wideband} require accurate calculations of surface impedance and kinetic inductance to predict gain and bandwidth; yet, 3D FEA-assisted modeling techniques have not been extensively explored.
The simulated transmission spectrum of the superconducting devices in this work demonstrates close agreement with the measured frequency responses and suggests a potential approach for studying KI-TWPAs solely based on their circuit layout and superconducting material properties.
An improved coupled mode theory~\cite{haus1991coupled}, employing the superconductor modeling techniques presented in this work, could pave the way for the realization of wideband, high-gain, quantum-limited noise KI-TWPAs with minimized ripples.

Despite the systematic improvements in predicting Hamiltonian over conventional methods, significant discrepancies in the predicted dispersive shifts of certain qubits remain elusive.
To further improve the accuracy of the proposed method, a better understanding of electromagnetic properties of superconductors and nonlinear contribution of Josephson junction is essential.
Recent studies on the quasiparticles~\cite{McEwen2022resolving, iaia2022phonon} have shown that nonequilibrium quasiparticles are excessively generated in superconducting circuits, resulting in correlated quantum errors. 
The contribution of quasiparticles to the surface impedance, calculated in this work, is limited to thermally excited quasiparticles and needs to be extended to include nonequilibrium quasiparticles.
Furthermore, a recent study on the observation of Josephson junction harmonics~\cite{Willsch2024observation} suggests that the cosine potential approximation in this work should be replaced with a more rigorous model.
We anticipate that further refinement in modeling realistic superconductors and Josephson junctions, combined with the proposed circuit quantization method, will enable a more accurate circuit quantization method with minimal simulated error.
Thus, combining these attributes, the proposed quantization method incorporating the anomalous electromagnetic behavior of superconductors facilitates precise engineering of large-scale superconducting circuits for quantum information processing.

\section*{Methods}\label{sec_methods}
\subsection*{Calculation of Superconducting Electromagnetic Properties}\label{sec_method_simul}
The complex conductivity $\sigma_{s}$ of isotropic superconductors at temperature $T$ and frequency $\omega$ can be calculated as~\cite{zimmermann1991optical}
\begin{equation}
    \sigma_s(\omega, T) = \frac{i\sigma_n}{2\omega\tau}\left(\int_{\Delta}^{\omega+\Delta}I_1 \,dE+\int_{\Delta}^{\infty}I_2 \,dE\right),
\end{equation}
\begin{equation}
    \begin{aligned}
    I_1 =\tanh{\left(\frac{E}{2k_BT}\right)}&\biggl\{\left[1-\frac{\Delta^2+E(E-\omega)}{P_2P_4}\right]\frac{1}{P_4+P_2+i/\tau}\\
    &-\left[1+\frac{\Delta^2+E(E-\omega)}{P_2P_4}\right]\frac{1}{P_4-P_2+i/\tau}\biggr\} ,
    \end{aligned}
\end{equation}
\begin{equation}
    \begin{aligned}
    I_2 =\tanh{\left(\frac{E+\omega}{2k_BT}\right)}&\biggl\{\left[1+\frac{\Delta^2+E(E+\omega)}{P_1P_2}\right]\frac{1}{P_1-P_2+i/\tau}\\
    &-\left[1-\frac{\Delta^2+E(E+\omega)}{P_1P_2}\right]\frac{1}{-P_1-P_2+i/\tau}\biggr\}\\
    +\tanh{\left(\frac{E}{2k_BT}\right)}&\biggl\{\left[1-\frac{\Delta^2+E(E+\omega)}{P_1P_2}\right]\frac{1}{P_1+P_2+i/\tau}\\
    &-\left[1+\frac{\Delta^2+E(E+\omega)}{P_1P_2}\right]\frac{1}{P_1-P_2+i/\tau}\biggr\},
    \end{aligned}
\end{equation}
where Boltzmann's constant $k_B$, $P_1=\sqrt{(E+\omega)^2-\Delta^2}$, $P_2=\sqrt{E^2-\Delta^2}$, $P_3=\sqrt{(E-\omega)^2-\Delta^2}$, and $P_4=i\sqrt{\Delta^2-(E-\omega)^2}$.
Here, $\tau$ is the electron relaxation time, defined as $\tau=l/v_F$, while $v_F=\pi\xi\Delta_0/\hbar$ is the Fermi velocity~\cite{tinkham2004introduction} and $l=\pi\mu_0\Delta_0\lambda_L^2\xi\sigma_n/\hbar$ is the electron mean free path~\cite{linden1994modified}.
Based on the FORTRAN program for calculating $\sigma_{s}$ in Ref.~\cite{zimmermann1991optical}, we implement an in-house Python program with automatic conversion between superconducting properties of $l$, $v_F$, $\Delta_0$, and $\sigma_n$.
We assume the frequency is much smaller than the superconducting gap energy, which can be converted into a frequency unit.

At specific $T$ and $\omega$, the surface impedance $Z_s$ of a bulk superconductor is defined as
\begin{equation}
    Z_{s,\textrm{bulk}}(\omega, T) = \sqrt{\frac{i\mu_0\omega}{\sigma_s}},
\end{equation}
where $\mu_0$ is the vacuum permeability.
For a superconducting film with a thickness $d$, a simple thin film approximation formula has been derived by A. Kerr~\cite{kerr1999surface} as
\begin{equation}
    Z_{s,\textrm{film}}(\omega, T) = \sqrt{\frac{i\mu_0\omega}{\sigma_s}}\coth{\left(\sqrt{i\mu_0\omega\sigma_s}d\right)}.~\label{eq_appendix_Zs_film_full}
\end{equation}

The two-fluid model describes the real $\sigma_1$ and imaginary $\sigma_2$ parts as the currents carried by the quasiparticles and Cooper-pairs in a superconductor, respectively.
In theory, at $T\ll T_c$, the real part of $\sigma_s$ is negligibly smaller than the imaginary part of $\sigma_s$ allowing the assumption $\sigma_s\approx-i\sigma_2$.
We note that while the two-fluid model predicts vanishing of the surface resistance of a superconductor at $T=0~$K~\cite{bardeen1957theory}, the residual $R_s$ values have been measured in various superconductors~\cite{oates1991surface, zhou1992analytical, posen2020ultralow}.
However, the level of dissipation caused by the residual $R_s$ is not comparable to other major coherence loss mechanisms, such as dielectric two-level system loss~\cite{Lisenfeld2015} or spontaneous emission loss~\cite{park2024characterization}, and the dissipative real part of $Z_s$ can be ignored.
Thus, Eq.\ref{eq_appendix_Zs_film_full} can be simplified to $Z_{s,\textrm{film}}\approx i\sqrt{\mu_0\omega/\sigma_2}\coth{(\sqrt{\mu_0\omega\sigma_2}d)}$, yielding the same result in Eq.~\ref{eq_Zf} in the main text.

In order to calculate $Z_s$ of superconducting niobium films in this work, we assume that $\lambda_L$ and $\xi$ of our niobium film are 33.3~nm and 39~nm, respectively, at $T=10~$mK as typical values~\cite{gao2008physics}.
The ratio of the superconducting gap energy to the critical temperature, $\Delta_0/(k_B T_c)$, is assumed to be 1.76, consistent with that of weakly coupled BCS superconductors~\cite{tinkham2004introduction}.

\subsection*{Niobium deposition and lithography}\label{sec_method_fab}
The devices used in this work were fabricated on 550 $\mu$m-thick c-plane sapphire substrate.
The wafer was precleaned by stirring in N-methyl-2-pyrrolidone, followed by sonication in acetone and isopropanol, and finally dried with nitrogen blowing.

The niobium thin film was deposited on the sapphire substrate using DC magnetron sputter (power 1~000 W, pressure 6 mTorr) for 35 seconds at a rate of 1 nm/s.
Prior to photolithography, alignkeys for electron beam lithography and chip dicing were fabricated.
The S1813 photoresist was coated with 4000 RPM and baked for 1 min 115~$^{\circ}$C.
The patterns were developed in AZ 300 MIF developer, followed by rinsing in deionized water.
Then, the niobium was etched using SF$_6$ gas for 20~s.
The residual photoresist was removed by rinsing in AZ 300 MIF again and O$_2$ plasma cleaning for 5~min.

\subsection*{Measurement Setup}\label{sec_method_measure}
The devices used in this work were mounted to the shielded mixing chamber plate of a dilution refrigerator, at the base temperature of about 10 mK.
For qubit XY control, pulse sequences were synthesized by using an arbitrary waveform generator (Tektronix AWG5014C) and an analog signal generator (Agilent E8257D).
We employed a vector network analyzer~(Agilent E5071C) for readout resonator spectroscopy.
The XY control lines and readout lines were attenuated by a total of 60 dB and 90 dB, respectively, to suppress thermal noise.
The output line for readout was amplified by a cryogenic high electron mobility transistor (HEMT) amplifier, followed by a series of amplifiers at room temperature.

\subsection*{Supplementary information}
See the supplementary material for additional details on the circuit quantization methods and FEA simulation techniques.

\subsection*{Acknowledgements}
This research was supported by National R\&D Program through the National Research Foundation of Korea~(NRF) funded by Ministry of Science and ICT~(2022M3I9A1072846) and in part by the Applied Superconductivity Center, Electric Power Research Institute of Seoul National University. This work was also partly supported by the New Faculty Startup Fund from Seoul National University. This work was also supported by the National Research Foundation of Korea~(NRF) grant funded by the Korea government~(MSIT) (No. RS-2024-00413957) and Quantum Computing Research Infrastructure Construction~(NRF-2019M3E4A1079894 and NRF-2022M3K2A1083855).

\section*{Declarations}
\subsection*{Conflict of interest/Competing interests}
The authors have no conflicts to disclose.

\subsection*{Author contribution}
S.H.P. conceived and developed the methods, and performed the numerical simulations.
G.C., G.P., and J.C. fabricated the sample devices and performed the experiments.
E.K. verified the methods and analyzed the results.
J.C. and Y.C. designed the devices.
Y.-H.L. supervised and guided the experiments during the project.
S.H. contributed to the methods and supervised the project.
S.H.P. wrote the manuscript with feedbacks from all authors.

\subsection*{Data availability}
All relevant data supporting the document are available upon reasonable request.

\subsection*{Code availability}
All relevant data supporting the document are available upon reasonable request.


%
  
\end{document}